\documentstyle[12pt]{article}
\def\header{\begin{flushleft}
            DESY 94-089\\ZU-TH 14/94\\June 1994
            \end{flushleft}}
\def\eq{eq.~}

\columnwidth\textwidth

\def\Im{\mbox{Im}}
\def\acp{a_{CP}}
\def\GSW_sign{}
\newcommand{\qsl}{q \hspace{-6pt} / \hspace{4pt}}
\newcommand{\ba}{\begin{array}}
\newcommand{\ea}{\end{array}}
\newcommand{\bd}{\begin{displaymath}}
\newcommand{\ed}{\end{displaymath}}
\newcommand{\be}{\begin{equation}}
\newcommand{\ee}{\end{equation}}
\newcommand{\bea}{\begin{eqnarray}}
\newcommand{\eea}{\end{eqnarray}}

\def\ub{\bar{u}}

\def\cb{\bar{c}}
\def\sb{\bar{s}}

\def\bra{\langle}
\def\ket{\rangle}

\def\a{\alpha}
\def\b{\beta}
\def\g{\gamma}
\def\d{\delta}

\def\L{\Lambda}

\def\to{\rightarrow}
\def\pol{\epsilon}
\begin{document}
\thispagestyle{empty}
\header \vspace*{2cm} \centerline{\Large\bf Branching ratio and direct
  CP-violating } \centerline{\Large\bf rate asymmetry of the rare
  decays }
\centerline{\Large\bf B$\to K^\ast\bf\gamma$ and
  B$\to\bf\rho\bf\gamma$ $\footnote{partially supported by
    Schweizerischer Nationalfonds.}$} \vspace*{2.0cm}
\centerline{\large C. Greub$^a$, H. Simma$^b$, D. Wyler$^a$}
\vspace*{0.5cm} \centerline{\large $^a$ Universit\"at Z\"urich, 8057
  Z\"urich, Switzerland} \vspace*{0.5cm} \centerline{\large $^b$
  Deutsches Elektronen Synchrotron DESY, 22603 Hamburg, Germany}
\vspace*{3cm} \centerline{\Large\bf Abstract} \vspace*{1cm} We
calculate CP-violating rate asymmetries in the rare radiative decays
$B^\pm\rightarrow K^{\ast\pm} \gamma$ and $B^\pm\rightarrow \rho^\pm
\gamma$.  They arise because of the interference between
leading-order penguin amplitudes and one-gluon corrections with
absorptive phases, and provide unambiguous evidence for direct
CP violation.  Complementing earlier studies, we
also investigate gluon exchange with the `spectator' quark. The bound
state effects in the exclusive matrix elements are taken into account
by a covariant model, which yields a branching ratio
$BR(B\rightarrow K^\ast \gamma) = (4-5)\times 10^{-5}$ in good
agreement with the observed value. The  bound state
effects increase the CP asymmetry, which is of order $1 \%$ in the
channel $B\rightarrow K^\ast \gamma$ and $15 \%$ for
$B \to \rho \gamma$.

\newpage
\subsection{Introduction}
There is much hope to detect CP violation in B-meson physics
\cite{JARLSK,SST,NQ}.
It would complement existing results in Kaon decays and help solving the
origin of this phenomenon. In fact, the large variety of decay modes
of the B-meson makes it possible to investigate CP violation from many
angles.

A few decay modes, such as $B^0\to K_S J/\Psi$, allow for a simple,
essentially model-free prediction of the CP-violating rate asymmetries
\cite{BS}.  These originate mainly from the mixing of $B^0$ and
$\overline{B^0}$, and are analogous to the $\epsilon$-parameter in
K-decay. However, one would also like to test direct CP violation in
the decay amplitudes, such as manifested by the $\epsilon'$ parameter
(for a recent review of direct CP violation, see \cite{WW}).
Theoretical investigations are restricted to inclusive decay modes,
which are difficult to relate to experimental data or require a
calculation of an exclusive decay within a phenomenological model.

It is well known that a non-vanishing asymmetry due to direct
CP-violation requires two interfering amplitudes with different weak
(CKM) and strong (rescattering) phases.  The main difficulty is to
calculate the latter. Within the present understanding of the basic
CP-violating mechanism, consisting in the simultaneous contribution of
several generations of particles to the physical processes in
question, it is natural to locate the strong phases in (penguin)
diagrams on the quark level, involving virtual heavy and on-shell
light particles of all these generations. This picture was introduced
a long time ago by Bander, Silverman and Soni \cite{BSS} and underlies
most treatments of direct CP violation in charged B-decays; we will
essentially follow these lines as well.

A prominent (penguin) decay is $B\to K^\ast \gamma$, recently observed
by the CLEO group with a branching ratio of $(4.5 \pm 1.0 \pm 0.9)
\times 10^{-5}$ \cite{CLEO}.  With only one hadron in the final state
it is an interesting candidate: One may hope that the theoretical
uncertainties of the calculated rate asymmetry
\be
\label{asymmgeneral}
\acp \equiv \frac{\Gamma(B^- \to K^{*-} \gamma)-
\Gamma(B^+ \to K^{*+} \gamma)}{
\Gamma(B^- \to K^{*-} \gamma)+
\Gamma(B^+ \to K^{*+} \gamma)}
\ee
are minimal and that the clear signature should make
experimental detection feasible.

While the rate asymmetry for the underlying quark process, $b\to s
\gamma$, vanishes up to the one-loop level (as the penguin diagram
generates an absorptive phase only when the photon is off shell),
Soares \cite{Soares} showed that there is a non-zero effect at two
loops.  When true hadronic transitions are considered, also bound
state effects must be included. They can give rise to additional
contributions due to absorptive parts of diagrams where additional
gluons are emitted from inside of the penguin-loop and couple to
the other constituents of the hadrons. In this paper we include
these effects and estimate the rate asymmetry in charged B-decays
using a simple model for the mesons.

\subsection{Perturbative contributions}
We write the effective Hamiltonian for the decay $B\to K^\ast \gamma$ as
\bea
{\cal H}_{eff} & = &  \frac{4 G_{F}}{\sqrt{2}} \, v_c \left(
C_2(\mu) O_2^{(c)} + C_7(\mu) O_7 + ... \right) \\
& + &  ( c \to u)\,. \nonumber
\label{Heff}
\eea
where
\be
\label{O_2}
O_2^{(q)} =
(\bar{q}_{\alpha} \gamma^\mu L b_{\alpha}) \cdot
(\bar{s}_\beta \gamma_\mu L q_{\beta}) \,
\ee
is the familiar four-Fermi operator and
\be
O_7 = \frac{e}{16\pi^2} \, \bar{s}_{} \, \sigma^{\mu
\nu} \, (m_b  R + m_s  L) \, b_{} \ F_{\mu\nu}
\label{O_7}
\ee is the magnetic dipole operator.  $L$ and $R$ are the left and
right projectors.  The operator $O_7$ arises from diagrams with an
internal top-quark, and hence its coefficient is proportional to $v_t
\equiv V_{tb} V_{ts}^\ast$.  It is convenient to express $v_t$ in terms
of the two independent CKM combinations $v_i \equiv V_{ib} V_{is}^\ast$
($i=u,c$) by making use of the unitarity relation $v_t = -v_c - v_u$.
This enables us to write the amplitudes \footnote{Of course, there are
several amplitudes corresponding to the different helicity states of
$K^\ast$ and $\gamma$. However, as long as not both the polarization of
the photon and the angular distribution of the decay products of the
$K^\ast$ are observed, there are no other CP-violating observables
besides the
rate asymmetry and we ignore the helicity structure throughout.}
for our process in the convenient form
\be
\label{amp}
A = v_u A_u + v_c A_c\ .  \ee For the decay B$\to \rho\gamma$ one
simply replaces the $s$-quark in the above expressions for the
operators and in the CKM factors by the $d$-quark.  The ellipses in
\eq(\ref{Heff}) denote other operators; in particular, four-Fermi
operators whose one-loop matrix elements in general contribute to the
rate. As pointed out in \cite{MISI}, their effects can, at least in
leading logarithmic approximation, be absorbed into an effective
coefficient $C_7^{eff}(\mu)$. For a top-quark mass of $174$ GeV,
the corresponding values of the Wilson coefficients are given by
$C_2(5 \ GeV) = 1.096$ and $C_7^{eff}(5 \ GeV) = - 0.305  \GSW_sign$
\footnote{
  The relative sign between $C_2$ and $C_7$ depends on the
  definition of the covariant derivative; we use
  $D = \partial + \GSW_sign i g_s T^a A^a  + \GSW_sign i e Q A$}
\cite{MISI,PIMA};
these values include some next-to-leading log (NLL) corrections.
Of course, a systematic calculation at the NLL
level not only requires the knowledge of the coefficients to this
precision, but also two-loop QCD-corrections to the matrix elements
(real and imaginary parts) have to be calculated.

The CP violating asymmetry $\acp$ has the form
\cite{papi}
\be
\label{asy}
\acp = \frac{- 4 \Im[v_u v_c^\ast] \, \Im[A_u
A_c^\ast]} {|v_u A_u + v_c A_c|^2 + |v_u^\ast A_u + v_c^\ast A_c|^2} \ee
which implies the well known fact that the absorptive parts of $A_u$
and $A_c$ must be evaluated.  These are subleading, i.e. vanish in the
leading log approximation and when the matrix element are treated only
at tree level.  In fact, the imaginary parts do not arise from the
coefficients $C_i$, but only from the matrix elements of the various
operators when evaluated to sufficient order in $\alpha_s$.
Therefore, when we consider an expansion of the amplitude in
$\alpha_s$, it will be consistent to neglect subleading terms,
except in the imaginary parts in the numerator in eq.(\ref{asy}).
In particular, these absorptive parts are renormalization scheme
independent, which is of course not generally true for the real parts
of the one-loop matrix elements.
The leading contribution to the decay rate comes from the tree-level
matrix element of $O_7$ in Fig.~1 which does not involve the spectator
anti-quark.

At next order in $\alpha_s$ there are many terms. In particular, when
considering the bound state, one would also include, in a perturbative
calculation, gluon exchange between the $b$-quark or its decay
products and the spectator line.  In fact, in a particular treatment
of high momentum transfer processes \cite{BLS}, the gluon exchange
plays the major role. In this scheme the constituents have essentially
no transverse momentum which must be provided by the gluon.  However,
in B-meson decays this approach tends to yield too small branching
ratios \cite{BLS} and is likely to serve only as corrections to the
decay width.  To estimate their importance, we have analyzed some
higher order graphs in the framework of our model (see below) and
found that they only give corrections of up to order $15\%$ in the
rate.  We therefore expect that an expansion in $\a_s$ is
meaningful and that very soft gluons can be absorbed in a suitable
wave function. Thus we only include the graph of Fig.~1 to calculate
the rate.

However, as pointed out, the higher orders are essential for the
asymmetry and we must isolate the important ones.  At order
$\alpha_s^0$ only the operator $O_7$ contributes to the decay
(after absorbing the contributions of four-quark operators into
$C_7^{eff}$);
thus, $A_c$ and $A_u$ are equal and real at lowest order. Including
order $\alpha_s$ corrections, we write
$A_c = A^0 + \alpha_s A_c^1$ and $A_u = A^0 + \alpha_s A_u^1$.
Then, $\Im[A_u A_c^\ast]$ in eq.(\ref{asy}) reads
\be\label{imampl} \Im[A_u A_c^\ast] =
\alpha_s A^0 \left( \Im A_u^1 - \Im A_c^1 \right) \quad .
\ee
This shows explicitly that the internal particles, which generate the
absorptive phase, must have different masses for a nonzero asymmetry
(i.e. the $i=u,c$ quarks from $O_2^{(i)}$ must be involved in $\Im A_i$).
All operators omitted in \eq(\ref{Heff}) enter with a coefficient
proportional to $v_u+v_c$ and therefore do not yield an absorptive
phase which contributes to the asymmetry at order $\a_s$.
For the same reason, absorptive parts of matrix
elements of the operators $O_7$ and $O_8$ (the analog to $O_7$, but
with a gluon instead of the photon) do not contribute; only the loop-level
matrix elements of $O_2$ with a gluon emitted from the loop are relevant.

There are basically two classes of diagrams depending on whether the
gluon from the internal quark loop is attached to the final $s$-quark
(Fig.~2) or to the spectator (Fig.~3).

The first class has already been studied in ref.~\cite{Soares} and
corresponds to the finite part of those QCD radiative corrections which
are also responsible for the considerable operator mixing of $O_2$
into $O_7$ \cite{PIMA,MISI}. For completeness, and ease of use, we
will recast the result of Soares in the framework of effective operators
which also gives rather compact expressions.  Applying the language of
Cutkosky rules, the absorptive parts arise from ``cuts'' which contain
the $u\ub$ or $c\cb$ pair of $O_2$, respectively. Other cut diagrams
either vanish or would generate an ``universal'' phase, i.e. one multiplied
by $v_u+v_c$, which contributes only at higher order to the rate asymmetry.

The relevant cut diagrams are shown in Fig.~2. They
yield local contributions
proportional to the tree-level matrix element of $O_7$ (i.e. proportional
to the contribution of Fig.~1)
\be
\label{o2soares}
\Im \bra s\g|O_2^{(i)}|b\ket_{Fig.2a,b} = g_s^2 \, Q_{a,b} \,
C_N \, \rho_{a,b} (m_i,m_b) \,
\bra s\g|O_7|b\ket_{\rm tree}\, ,
\ee
where $C_N = (N^2-1)/(2N)$ is a color factor ($C_N = 4/3$ for $N=3$
colors). $Q_a = -1/3$ and $Q_b = 2/3$ are the charges of the quarks to
which the photon couples in Fig.~2a,b, respectively. The masses of the
internal and of the $b$-quark are denoted by  $m_i$ and $m_b$,
respectively. To evaluate the factors $\rho_{a,b}$ in \eq(\ref{o2soares}),
we apply the Cutkosky rules in the form
\be
\Im A(b\to f) = {(2\pi)^4\over 2} \sum_i \int
\prod_{n=1}^{N_i} {d^3 p_n\over (2\pi)^3 2E_n} \,
\d^4\Bigl( p_b - \sum_{n=1}^{N_i} p_n \Bigr)\,
\hat A(b\to i) \, \hat A(i\to f)
\quad,
\ee
where $\hat A$ are the (real) amplitudes of the corresponding
subprocesses and the summation runs over all
possible intermediate states $i$.

The diagrams with only the gluon coupled to the quark loop (Fig.~2a)
contain penguin subdiagrams with insertion of $O_2^{(i)}$. They can
be represented by an effective $\bar{s}bg$ vertex
\be \label{drodsky1}
I^a_\mu = \GSW_sign \frac{g_s}{16\pi^2} \,
V\Bigl(\frac{q^2}{m_i^2}\Bigr) \,
\left(q^2 \gamma_\mu - q_\mu \qsl \right) \, L \, \frac{\lambda^a}{2}
\quad , \ee
where $q$ is the momentum of the gluon, and $V$ is proportional
to the vacuum polarization with internal $i=u$ or $c$ quarks.
With this building block the sum of the diagrams in Fig.~2a yields
\be
\rho_a(m_i,m_b) = \GSW_sign \frac{1}{16 \pi^2} \, \frac{1}{m_b^6}
 \int_{4 m_i^2}^{m_b^2} \, \Im V\Bigl(\frac{q^2}{m_i^2}\Bigr) \,
(m_b^2- q^2)^2 \, dq^2 \quad ,
\label{rho_a}
\ee
where
\be
\label{vacuum}
\Im V(x) = \frac{2 \pi}{3} \,
\frac{x+2}{x} \sqrt{\frac{x-4}{x}} 
\, \, \Theta(x-4) \quad .
\ee

The sum of the diagrams with the gluon {\it and} the photon emitted
from the quark loop (Fig.~2b) involves an effective $\sb b g \gamma$
vertex which can be derived from ref.~\cite{SW}
\bea
\label{drodsky2}
I^a_{\mu \nu} &=& -\frac{g_s e Q_u}{8 \pi^2} \,
\frac{\lambda^a}{2} \,
\left\{i \epsilon_{\beta \mu \nu \alpha} \,
(q^\beta \Delta i_5 + p_\gamma^\beta \Delta i_6) \ba{l} ~\\~\ea \right.
\nonumber \\
&& \hspace{0.5cm} \left.
 + i \frac{\epsilon_{\rho \sigma \mu \alpha}}{p_\gamma q} \,
 q^\rho p_\gamma^\sigma q_\nu \Delta i_{23} +
 i \frac{\epsilon_{\rho \sigma \nu \alpha}}{p_\gamma q} \,
 q^\rho p_\gamma^\sigma p_{\gamma \mu} \Delta i_{26} \right\} \,
\gamma^\alpha \, L  \quad .
\eea
In general, for the gluon and the photon being off shell,
the quantities $\Delta i_n \equiv \Delta i_n(z_0,z_1,z_2)$
are functions of the three variables $z_0 = s/m_i^2$,
$z_1 = q^2/m_i^2$ and $z_2 = p_\gamma^2/m_i^2$,  where
$q$ and $p_\gamma$ denote the four-momenta of the gluon and the
photon, respectively, and $s$ is the invariant mass squared
of the internal quark pair.
In the present situation, where the photon is on shell, the
form factors $\Delta i_n$ can be written as
\begin{eqnarray}
\label{deltaiexpl}
\Delta i_5(z_0,z_1,0) &=& -1 + \frac{z_1}{z_0-z_1} (Q_0(z_0) - Q_0(z_1))
-\frac{2}{z_0-z_1} (Q_-(z_0)-Q_-(z_1))  \nonumber \\
\Delta i_6(z_0,z_1,0) &=& +1 + \frac{z_1}{z_0-z_1} (Q_0(z_0) - Q_0(z_1))
+\frac{2}{z_0-z_1} (Q_-(z_0)-Q_-(z_1))  \nonumber \\
\Delta i_{23}(z_0,z_1,0) &=& \Delta i_5(z_0,z_1,0)
= - \Delta i_{26}(z_0,z_1,0) \quad .
\end{eqnarray}
The functions $Q_0$ and $Q_-$ are defined by the integrals
\be
\label{q0qmint}
Q_-(x) = \int_0^1 \, \frac{du}{u} \,
\log \left( 1-xu(1-u) \right)
\quad , \quad
Q_0(x) = \int_0^1 \, du \log \left( 1-xu(1-u) \right) \quad ,
\ee
and have imaginary parts for $x\ge 4$. On the sheet $\Im x \ge 0$
they are given by
\be
\label{q0qm}
\Im Q_-(x) = - 2 \pi \log \left(\frac{\sqrt{x} + \sqrt{x-4}}{2} \right)
\quad , \quad
\Im Q_0(x) = - \pi \sqrt{\frac{x-4}{x}}\ .
\ee
In terms of these functions, we obtain for the total contribution of the
diagrams in Fig.~2b
\bea
\rho_b(m_i,m_b) & = & \GSW_sign \frac{1}{8 \pi^2} \,
\int \, d \hat{E}'_s \, dz \, \frac{\hat E'_s}{[1-\hat E'_s(1-z)]^2}
\nonumber \\
&& \times \left\{ \left( 2 - 4 \hat E'_s (1-z)\right) \,
   \frac{m_i^2}{m_b^2}\,  \Im Q_-\Bigl(\frac{s}{m_i^2}\Bigr)
   \right. \nonumber \\[2mm]
&& + \left. \left( 3 \hat E'_s - \hat E'_s z
   -4 \hat E^{'2}_s (1-z) \right) \Im Q_0\Bigl(\frac{s}{m_i^2}\Bigr) \right\}
   \quad , \label{rho_b}
\eea where $\hat E'_s m_b$ is the energy of the intermediate (cut)
$s$-quark in the rest frame of the decaying $b$-quark and $z$ is the
cosine of the angle between the intermediate $s$-quark and the photon.
In terms of the integration variables the invariant mass squared
of the internal quark pair reads $s = m_b^2 (1 - 2 \hat{E}'_s)$.

At $z=-1$ the integral in eq. (\ref{rho_b}) is divergent.  In an
inclusive (free) $b$-quark decay this divergence would have to be
cancelled by including the radiation of additional gluons or
$s\sb$-pairs.  In the present exclusive situation this logarithmic
divergence is cut off by requiring that the momentum squared $q^2$ of
the virtual gluon is sufficiently off shell in order to avoid double
counting with contributions from the wave functions. The condition
$q^2 = -m_b^2 \hat{E}'_s (1+z) \le - \Lambda^2$ yields the integration
intervals
\begin{eqnarray}
\label{intervals}
\frac{\Lambda^2}{2m_b^2} \le & \hat{E}'_s & \le
\frac{1}{2} \nonumber \\
-1+\frac{\Lambda^2}{m_b^2 \hat{E}'_s} \le & z & \le 1 \quad .
\end{eqnarray}

We close this section with a physical picture of the diagrams
involved.  For instance, when the internal quark loop contains a
$c\cb$ pair, the above diagrams can be viewed as perturbative
contributions to the decay of the B-meson into some $(s\cb)$ and
$(c\ub)$ states, e.g. a $D_s^{(\ast)}$ and a $D^{(\ast)}$ meson. These
intermediate hadrons rescatter then into the final state $K^\ast +
\gamma$ by radiating the photon and exchanging gluons in all possible
ways. This implies the presence of the diagrams (see Fig.~3) of the
second class, involving the spectator line. They yield non-local
contributions and their explicit form depends on the model by which
the meson states are described; the one we use will be presented in
the next section.

\subsection{A model for the exclusive matrix elements}
To describe exclusive decays of $B$ mesons involving a light meson in
the final state, like $B \to \pi e \nu$ and $B \to \rho e \nu$, we have
developed a covariant model \cite{GW} which yields good results for
the branching ratio and energy spectrum in these decays; we apply
it here to the rare decay $B \to V \gamma$, where $V = K^\ast$ or $\rho$.
In fact, since most models coincide when
the hadronic momentum transfer
squared is approximately zero, we expect that the results
are relatively model independent. We assume that both $B$ and
$K^\ast$ can be described by two (effective) constituents only:
$B=(b,\bar{q})$ and $K^\ast= (s,\bar{q})$. To guarantee covariance of
the results we require that the four-momenta of the constituents add
up to the four-momentum of the bound state.  Due to binding effects,
this condition can only be fulfilled if at least one of the two
constituents has a variable mass \footnote{This is quite similar in
the model of Altarelli et al.  \cite{Altarelli}}.  We take the mass $m_q$
of the spectator anti-quark to be fixed; more specifically we put $m_q=0$
(for reasons which are related to our prescription of the light mesons,
see below). Consequently, the $b$-quark mass becomes momentum dependent
\be \label{mbvar}
m_b = \sqrt{m_B^2 - 2 p_B p_q} \quad ,
\ee
where $p_B$ and $p_q$ are the four-momenta of the
$B$ meson and the spectator, respectively, and $m_B$ is the mass
of the $B$ meson.
Eq.~(\ref{mbvar}) only makes sense if the momentum $|\vec{p}_q|$ in the
$B$ rest-frame is restricted to $|\vec{p}_q| \le m_B/2$.
As derived in detail in ref. \cite{GW}, the $B$ meson can be
represented by a matrix $\Psi_B$ (in spinor and color space)
\begin{equation}
\label{matb}
\Psi_B = C_B \, \int \frac{d^3p_q}{2E_q (2\pi)^3} \,
\sqrt{\frac{m_B}{2(m_B^2 - p_B p_q)}}
\, \phi_B(|p_B p_q/m_B|) \,
\Sigma_B \otimes \, \frac{1}{\sqrt{N}} {\bf 1}_N \quad ,
\end{equation}
with
\begin{equation}
\label{pbhut}
\Sigma_B = -(\not\!p_b + m_b) \, \gamma_5 \, \not\!p_q \quad .
\end{equation}
For the wave function $\phi_B(p)$ we use a harmonic Ansatz
\begin{equation}
\label{fwave}
\phi_B(p) =
\exp\left\{-p^2/(2 p_F^2)\right\} \quad .
\end{equation}
The normalization factor $C_B$ is chosen such that
$\int d^3p_q |C_B \Phi_B(|\vec{p}_q|)|^2 = (2\pi)^3$
when the integration region is restricted to $|\vec{p}_q| \le m_B/2$.
Numerically, $C_B \approx \sqrt{8} \, \pi^{3/4} \, p_F^{-3/2}$
to a very high precision.  The only parameter of the wave function,
$p_F$, is fixed in such a way that our model yields the right value
for the meson decay constant $f_B$. Numerical values of $p_F$
for different choices of $f_B$ are given in table~1.

While the above construction is adequate for the $B$ meson, it turns
out to be impossible to describe the light (vector) meson
along these lines because whatever one chooses for the
wave function, it is impossible to get a value for the decay constant
which is compatible with experimental data. (Here we refer to the
analogous statement given in detail for the pion in ref.
\cite{GW}). Therefore, we use a different picture for the
light meson, assuming that the meson is represented by two constituents
with parallel momenta $p_s = y p_V$ and $p_q = (1-y) p_V$, respectively.
As the vector mesons considered here are much lighter than the $B$ meson,
we neglect their masses and consequently also those of the constituents.
The final state vector meson is represented by the matrix
\begin{equation}
\label{matkstar}
\Psi_V = C_V \, \int_0^1 dy \, \phi_V(y) \, \Sigma_V
\otimes \, \frac{1}{\sqrt{N}} {\bf 1}_N \quad ,
\end{equation}
with
\be\label{pfhut}
\Sigma_V = \lim_{m_V \to 0} \not\!\pol_V^\ast (\not\!p_V + m_V) \ ,
\ee
where $p_V$ and $\epsilon_V$ are the momentum and polarization vector
of the meson, respectively.
$\phi_V(y)$ is the quark distribution amplitude in the vector meson,
and the normalization factor $C_V$ is determined by the value
of the decay constant $f_V$, defined as
$\bra 0 | \bar{u} \gamma_\mu s | V \ket = m_V \, 
\, f_V \, \epsilon_V^\mu$ ($f_{K^\ast} \approx f_\rho \approx 
216$ MeV).
Explicitly, we take
\be
\label{Phi_V}
\phi_V(y) = 6 y(1-y)\,(1+\cdots)  \quad ; \quad
C_V = 
f_V/(4 \sqrt{N}) \quad
\ee
where the ellipses denote deviations from the asymptotic form of the
wave function and $SU(3)_{\rm Flavour}$ breaking effects (see e.g.
ref.~\cite{CZreport}).

The exchange of extra gluons may lead to ambiguities (double counting
and IR divergences) in two situations: In the first, there is a quark
line, like the $b$-line in Fig.~4a, which is off shell because of the
exchange of a gluon which can be absorbed into one of the meson wave
functions.  In order to avoid counting this configuration a second time
in the perturbative treatment, we should impose a condition that the
quark line is sufficiently off shell.

In the second, gluons which couple directly to a constituent
quark of a meson should not be included when they are almost on shell:
This case would be degenerate with the contribution from the adequate
wave function for a three-particle Fock state (e.g. $\ub g s$
in Fig.~2); since the latter are not included in our simple model
treatment, we should not include such almost on-shell gluons as well.

Both conditions will be built into our model by requiring that quarks
or gluons in the above situations are off shell by at least an amount
$\Lambda$. (Of course, the cutoff parameter may a priori be different
in the two situations; however, for simplicity we take the same value
$\L$ in both cases.) This parameter is part of the definition of our model
and a value of $\L = 0.2 - 1 $ GeV should be physically reasonable.
This automatically cuts off possible soft and collinear divergencies.

In the following, we denote by  $p_\gamma$ ($p_V$) and $\pol_\gamma$
($\pol_V$) the momentum and polarization of the photon (vector meson),
respectively. The transition matrix elements of the effective operators
in \eq(\ref{Heff}) have the general form (for an on-shell photon)
\be
\label{me_def}
\bra V \gamma | O_i | B \ket \equiv
e \, m_B \,\pol_\gamma^\mu \left\{ i \epsilon_{\mu\nu\a\b}
p_\gamma^\nu \pol_V^\a p_V^\b \,F_1[O_i]
+ \Bigl((p_\gamma \pol_V)\, p_{V\mu} - (p_\gamma p_V)\, \pol_{V\mu}\Bigr)
\, F_5[O_i] \right\}
\ee
which serves as a definition for the `form factors' $F_{1,5}[O_i]$.
In the following, when the mass of the final-state meson is neglected,
we have always $F_1=F_5$ due to the $\sigma_{\mu\nu}(1+\g_5)$ structure
of $O_7$; hence, we shall drop the subscript of $F$ \footnote{
   For the operator $O_7$, our definition of $F$ coincides with the
   form factor $2F_1$ commonly used in the literature}.

In the framework of our model for the bound states, the hadronic matix
elements are determined by
\bea \label{me_trace}
\bra V \gamma | O_i | B \ket & = &  C_B \, C_V \, \int \,
   \frac{d^3p_q \, dy}{(2\pi)^3 2 E_q} \,
   \sqrt{\frac{m_B}{2(m_B^2 - p_B p_q)}} \nonumber \\
&& \times \, \phi_B\Bigl(\frac{p_B p_q}{m_B}\Bigr) \, \phi_V(y) \,
   \frac{1}{N} \mbox{Tr} \left[
   \Sigma_V \, M_{sb} \, \Sigma_B \, M_{qq'} \right] \quad ,
\eea
where the trace is in Dirac and color space.
The spin projectors $\Sigma_B$ and $\Sigma_V$ in \eq(\ref{me_trace})
are given in eqs.~(\ref{pbhut}) and
(\ref{pfhut}), respectively, and the matrices $M_{sb}$ and $M_{q'q}$
(acting in spinor and color space) are related to the
quark-level matrix elements by
\be
\bra s \bar{q}' \gamma| O_i | b \bar{q} \ket = \bar{u}_s \, M_{sb} u_b
\cdot \bar{v}_q M_{qq'} v_{q'} \quad .
\ee
Note, that in the case of a transition where the spectator is not directly
involved, $M_{qq'}$ can be written as \cite{GW}
\be \label{mqq}
M_{qq'} = 2 \gamma_0 \, (2 \pi)^3 \, \delta^3(p_q - p_{q'})
\otimes {\bf 1}_N \quad .
\ee

{}From the decomposition in \eq(\ref{me_def}) one readily derives the
decay width
\be
\label{widthgeneral}
\Gamma(B \to V\gamma) =
\frac{\a_{em} }{4} \, m_B^5 \, \vert F[{\cal H}_{eff}]\vert^2 \quad .
\ee
In the (leading) approximation when only the matrix element of $O_7$
is taken into account, one obtains
\be
\label{widthlowest}
\Gamma(B \to V \gamma) \approx
2\, G_F^2 m_B^5 \alpha_{em} \,
|v_c + v_u|^2 \, (C_7^{eff})^2 \, |F[O_7]|^2 \quad .
\ee

\subsection{Evaluation of $\bf F[O_7]$ and branching ratios}
As mentioned in section~2, only
$F[O_7]$ contributes to $\bra V \gamma |
{\cal H}_{eff} | B \ket $ in a leading-log calculation of the decay rate.
Evaluating the trace in \eq(\ref{me_trace}) and working out the $d^3p_q$
integration by making use of the 3-dimensional delta function (which
directly relates $m_b=m_B \sqrt{y}$  to the momentum fraction $y$)
we get
\be
\label{sigma7}
F[O_7] = - \frac{ \sqrt{m_B} \, C_B \, C_V }{4 \pi^2} \,
\int \, dy  \, \frac{y}{\sqrt{1+y}} \, \phi_V(y) \,
\phi_B\Bigl(\frac{m_B (1-y)}{2}\Bigr)  + {\cal O}(\a_s) \quad .
\ee

In order to get a rough idea of the effect of higher order contributions
to the amplitude for $B \to K^\ast \gamma$, we consider those order
$\a_s$ corrections to the matix element of $O_7$ which involve an extra
gluon exchange with the spectator. The corresponding diagrams, with the
gluon radiated either off the $b$- or $s$-quark, are shown in Fig.~4.
The comparison of the lowest order contributions in our model with these
diagrams is of particular interest, because the latter would be the
leading contributions in the approach of ref.~\cite{BLS}, where the
exchange of hard gluons is essential.

While the second diagram of Fig.~4 vanishes for $m_s=0$, the first
diagram yields the following contribution to $F[O_7]$
\bea \label{sigma71}
F[O_7]_{Fig.4} &=& \frac{g_s^2 }{\pi^2} \, m_B \, C_N
\, C_B \, C_V \nonumber \\
&\times& \int \, dy \, dE_q \, dz \, \phi_V(y) \, \phi_B(E_q) \,
\, \frac{ E_q^2 (m_B - 2 E_q) \, (2-y(1-z)) }
{N_b \, q^2 \, \sqrt{2(m_B - E_q)} }
\eea
where $E_q$ is the energy of the spectator in the rest frame \footnote{
  Since the model is covariant, the result is of course frame independent.}
of the B-meson, and $z$ is the cosine of the angle between the spectator
and the $K^\ast$-meson.
$N_b$ is the denominator of the $b$-quark propagator and $q^2$ is the
invariant mass squared of the exchanged gluon. In terms of the integration
variables we have
\be
\label{nbqq}
N_b = -(1-y) m_B^2 + 2 m_B E_q \quad , \quad
q^2 = -m_B E_q (1-y) (1-z) \quad .
\ee
{}From pure kinematics, both can become zero and generate divergencies
in \eq(\ref{sigma71}). However, this is avoided by the cutoffs included
in our model (see section~3): When the $b$-quark propagator, $N_b$,
goes to zero the corresponding gluon exchange can be viewed as already
been taken into account in an adequate wave function of the B meson.
We may therefore exclude these contributions in the perturbative treatment
and impose the cutoff $|N_b| \ge \Lambda^2$.

Similarly in $q^2$, a vanishing factor $(1-z)$ (the factor $(1-y) E_q$
is already cancelled by the wave function $\phi_V(y)$ and by $E_q$ in
the numerator) corresponds to the kinematical situation where the gluon
is on shell and parallel to the spectator. Such kinematical configurations
would also enter when we consider higher fock-components of the mesons,
e.g. $\ub g s$. Since we do not go beyond 2-Fock states here, we should
also cut off this kinematical region by imposing
the (covariant) condition $|q^2| \ge \Lambda^2$. Of course, this leads
to a weak (logarithmic) dependence of the results on the value of $\Lambda$.
We find that for $\Lambda = 0.2$ GeV (1.0 GeV) the non-leading corrections
of \eq(\ref{sigma71}) contribute at most 23\% (6\%) to the rate.

In order to estimate the branching ratio we normalize the rate,
evaluated at leading order, by the theoretical prediction for the
inclusive semileptonic decay width. To evaluate the latter, we apply
our bound-state model to the decaying B-meson and obtain
\be
\label{widthsemil}
\Gamma_{sl} =
\frac{G_F^2}{192 \pi^3} \, \vert V_{cb}\vert^2 \, C_B^2  \,
\int \, \frac{d^3p_q}{(2 \pi)^3} \, \vert \phi_B
(|\vec{p}_q|)\vert^2 \, \frac{m_b^6(p_q)}{m_B-E_q} \,
g\Bigl(\frac{m_c}{m_b(p_q)}\Bigr)
\quad ,
\ee
where $g(r) = 1 - 8r^2 + 8 r^6 -r^8 - 24 r^4 \, \log(r)$
is the usual phase space function.
As we work in the leading-log approximation we have neglected QCD
corrections in $\Gamma_{sl}$.

The branching ratio is then obtained by multiplying with the
measured value for $BR(B \to X \ell \nu_\ell)=0.11$:
\be
BR(B \to V \gamma) \equiv
\frac{\Gamma(B \to K^\ast \gamma)}{\Gamma_{sl}} \times 11 \%\ .
\ee

Finally, it is straigthforward to apply our description of the B-meson
to the inclusive decays $B\to X_s \gamma$. For the ratio of exclusive
to inclusive decay rates, which is the more convenient quantity
for phenomenological investigations, we obtain
\be
R_{K^\ast} \equiv \frac{\Gamma (B \to K^\ast + \gamma )}{
\Gamma (B\to X_s + \gamma )}
=\frac{m_{B}^5}{m_{b,eff}^5}\, 64\pi^4 \, \vert F[O_7] \vert^2\ ,
\ee
where $m_{b,eff}$ is given by
\be
\label{mbeff}
m_{b,eff}^5 \equiv \int \, \frac{d^3p_q}{(2 \pi)^3} \, C_B^2  \, \vert \phi_B
(|\vec{p}_q|)\vert^2 \, \frac{m_b(p_q)}{m_B-E_q} \, m_{b}^5(p_q) \quad .
\ee
Neglecting $SU(3)_{\rm Flavour}$ breaking effects, $R$ is the same
for $B \to K^\ast\gamma$ and $B \to\rho\gamma$.

\subsection{Absorptive parts from $\bf ImF[O_2]$ and rate asymmetries}
For the calculation of the $CP$ asymmetry the imaginary part
of $F[O_2]$ is needed. Corresponding to the contributions from the
different classes of diagrams (see Figures 2 and 3), we split it into
\be
F[O_2] = F[O_2]_{Fig.2a} + F[O_2]_{Fig.2b} + F[O_2]_{Fig.3a} +
F[O_2]_{Fig.3b} \quad .
\ee

The diagrams in Fig.~2, where {\it no} gluon is exchanged with the
spectator, are readily incorporated into our model.
As the effective structure of the relevant subdiagram is proportional
to $\bra s \gamma|O_7|b \ket$ [see \eq(\ref{o2soares})], the expression
is similar to $F[O_7]$ from \eq(\ref{sigma7})
\bea \label{soares1}
\Im F[O_2^{(i)}]_{Fig2a,b} & = &
- \sqrt{m_B} \frac{g_s^2}{4 \pi^2 } \, Q_{a,b}\, C_N \, C_B\,  C_V
\nonumber \\
& \times &
\int \, dy \, \frac{y}{\sqrt{1+y}} \, \phi_V(y) \,
\phi_B\Bigl(\frac{m_B (1-y)}{2}\Bigr) \, \rho_{a,b}(m_i,m_b) \quad ,
\eea
where $Q_a = -1/3$, $Q_b = 2/3$,  and $\rho_{a,b}$ have been defined
in section~2 [eqs. (\ref{rho_a}) and (\ref{rho_b})].
Note that $m_b = m_B \sqrt{y}$ depends on $y$ according to our
bound-state model; and therefore [see \eq(\ref{intervals})],
$\rho_b$ is only non-zero for
\be \label{yrange}
\frac{\Lambda^2}{m_B^2} \le y \le 1 \quad .
\ee

In the diagrams {\it with} gluon exchange to the spectator,
the photon can be emitted either from one of the external fermion
legs (Fig. 3a) or from the $i=u,c$ quark in the fermion loop (Fig. 3b).
Diagrams in which the photon is emitted from the $b$-
or $s$-quark line do not develop imaginary parts, because the
corresponding momentum squared of the gluon, $q^2$, is negative.
The diagrams in Fig.~3a where the photon is emitted from
the spectator quark involve the effective $\bar{s}bg$ vertex of
\eq(\ref{drodsky1}) and we obtain for their sum
\bea \label{brodsky1}
\Im F[O_2^{(i)}]_{Fig.3a} &=& \GSW_sign
   \frac{g_s^2}{16 \pi^4}\, \frac{1}{m_B^2}\,Q_u \,C_N\, C_B\, C_V
   \int \, dz \, dE_q \, dy \nonumber \\
&& \times \frac{E_q^2}{\sqrt{2(m_B - E_q)}} \,
   \phi_B(E_q) \, \frac{\phi_V(y)}{1-y} \,
   \Im V\Bigl(\frac{q^2}{m_i^2}\Bigr) \quad ,
\eea
where $\Im V$ is given in \eq(\ref{vacuum}) and the momentum of the
gluon is related to the integration variables according to
\be \label{qsquared}
q^2 = m_B^2 (1-y) - 2 m_B E_q + m_B E_q y (1-z) \quad.
\ee

In the diagrams of Fig.~3b the photon {\it and} the gluon is
emitted from the fermion loop, and the sum of the corresponding two
one-particle irreducible subdiagrams is represented by the effective
$\bar{s}bg\gamma$ vertex of \eq(\ref{drodsky2}).
After implementing it in the bound-state model,
we obtain in terms of the form factors $\Delta i_n$ of
\eq(\ref{deltaiexpl})
\begin{eqnarray}
\label{brodsky2}
\Im F[O_2^{(i)}]_{Fig.3b} &=&
\GSW_sign \frac{g_s^2 Q_u C_N C_B C_V}{64 \pi^4} \,
\int \, dz \, dE_q \, dy \,
\frac{E_q^2}{\sqrt{2(m_B - E_q)}} \,
\phi_B(E_q) \, \phi_V(y)          \nonumber \\
&& \times \, \left\{  \frac{4(1-z) E_q}{m_B} \, \Im \Delta i_5
 - 4 \Im \Delta i_6 + \frac{ 2 (1-z^2) E_q^2}{p_\gamma q}
                 \Im \Delta i_{23}  \right. \nonumber \\
&& \hspace{0.0cm}   \left.
 + \, \frac{ \left[m_B (1-y) - 2 E_q \right](1+z) m_B }{p_\gamma q}
 \Im \Delta i_{26} \right\}
\, \frac{1}{q^2}  \quad ,
%
\end{eqnarray}
where
\bea
q^2 & = & -m_B E_q (1-y) (1-z) \quad , \nonumber \\
p_\gamma q & = & \frac{m_B}{2}
[(1-y)m_B  - E_q (1+z)] \quad .
\eea
The arguments of the functions $\Delta i_n(z_0,z_1,0)$ are given by
$z_0 = (p_\gamma + q)^2/m_i^2$ and $z_1 = q^2/m_i^2$.
The cut on the virtuality of the gluon ($q^2 \le - \Lambda^2$)
leads to the restricted integration intervals
\bea
0 \le & y & \le 1 - \frac{\Lambda^2}{m_B^2} \nonumber \\
\frac{\Lambda^2}{2 m_B (1-y)} \le & E_q &
\le \frac{m_B}{2} \nonumber \\
-1 \le & z & \le 1 - \frac{\Lambda^2}{m_B E_q (1-y)}
\quad .
\eea

We can now evaluate the rate asymmetry $\acp$ defined in
\eq(\ref{asymmgeneral}). In terms of the above form factors
$F[O_k]$, the amplitudes $A_u$ and $A_c$ entering in \eq(\ref{asy})
are proportional to
\be
A_i \sim  C_7^{eff}(m_B) \, F[O_7] + C_2(m_B) F[O_2^{(i)}]\quad.
\ee
Retaining only the leading (order $\alpha_s$) terms we get
\be
\label{asyfinal}
\acp = -2 \frac{\Im[v_u v_c^\ast]}{|v_u+v_c|^2} \,
\frac{C_2(m_B)}{C_7^{eff}(m_B)} \,
\frac{\Im F[O_2^{(u)}] - \Im F[O_2^{(c)}]}{F[O_7]} \quad .
\ee
To render the dependence on the CKM matrix more transparent it is
convenient to use the Wolfenstein parametrization \cite{CKMW}
which yields
\be \label{wolf}
\frac{\Im[v_u v_c^\ast]}{|v_u+v_c|^2} =
\left\{
\ba{ll}
- \eta \lambda^2
&({\rm for\ } b\to s {\rm\ transitions})\\
\frac{\eta}{(1-\rho)^2+\eta^2} \ \ \
&({\rm for\ } b\to d {\rm\ transitions})\\
\ea
\right. \quad .
\ee

Our numerical calculation of the asymmetries is based on
a CKM matrix with $\lambda=0.2205$ [see \eq(\ref{wolf})]
and $\rho = -0.3$, $0.0$, and $+0.3$ for values of
$f_B = 150$ MeV, 200 MeV and 250 MeV, respectively.
We vary $\eta$ in the allowed range 0.15 \ldots 0.5 (for
$m_t=174$ GeV).
For the mass values we use $m_B=5.28$ GeV and $m_c=1.5$ GeV,
while $m_{b,eff}$ from \eq(\ref{mbeff}) is given in table~1
for different values of $f_B$.
The corresponding values of the CP asymmetries, together with
the branching ratio and $R$, are given in table~2 for the two
choices of the cutoff parameter $\Lambda= 1$ GeV and
$\Lambda = 0.2$ GeV.

In our approximation, $\alpha_s$ which we choose as $\alpha_s=0.2$,
enters linearly and the asymmetry is indirect proportional to
$C_7^{eff}(\mu)$ which suffers from a considerable uncertainty
related to the choice of $\mu$ \cite{AG,MISI}.
We have also estimated $SU(3)_{\rm Flavour}$ breaking effects in
the wave functions \eq(\ref{Phi_V}). Using a parametrization
as in ref.~\cite{ABS}, the branching ratio
for $B\to K^\ast\gamma$ ($B\to\rho\gamma$)
increases (decreases) by up to 30\%, compared to the
values in table~2, while $\vert \acp \vert$
decreases (increases) less than 4\%  (7\%), respectively.
We also note that for the contributions of the diagrams in Fig.~2
alone, the rate asymmetry is essentially independent of uncertainties
from the details of the hadronic matrix elements\footnote{
   In our model, the effect of the bound-state wave functions does
   not completely cancel because the argument $m_b$ of $\rho$ in
   \eq(\ref{soares1}) depends on the quark momenta.};
this is not the case for the diagrams involving the spectator
(see Fig.~3).

\subsection{Discussion}
In this paper we have calculated the rate and the CP-violating
asymmetry for B$\to K^\ast\gamma$. The branching
ratio of $4-5 \times 10^{-5}$ is in good agreement with recent CLEO
measurements \cite{CLEO} and other theoretical approaches \cite{AG,ABS}.
With a typical asymmetry near 1\%, about $10^{10}$ B-mesons are required
to resolve $\acp$ experimentally.
A simple rescaling yields for B$\to \rho\gamma$ an asymmetry of
above 10\% and a branching ratio of $10^{-6}$, requiring about $10^9$
B-mesons. Similar numbers of B-mesons are needed
for all such rate  asymmetries which test {\it direct} CP
violation. One expects such samples of B-mesons at the planned
hadronic B-facilities \cite{PS}.

The asymmetry receives contributions from pure spectator-type decays
(considered first in ref.~\cite{Soares}) {\it and} from transitions
involving the non-decaying meson constituents. The two add
constructively, a fact stressing the importance of bound state
effects and possible enhancements of the observables.

The occurence of both mechanisms is expected if one views the rate
asymmetry as arising from the rescattering of intermediate DD$_s$
or $K\pi$ meson pairs. In a quark picture, there is no reason to
neglect one of the possible gluon exchanges between the intermediate
$c$ or $u$ quark and the other constituents; taking into account both
contributions just reflects the symmetry of the problem.
If one neglects soft rescattering of the mesons, the asymmetry is
essentially a phase space effect (generated by the difference of the
$u$ and $c$ quark masses) and probably well represented by the quark
diagrams.

The asymmetry depends on a cutoff parameter for the momentum of the
exchanged gluon. Choosing it between  200 MeV and 1 GeV leads to a
variation of the asymmetry by about 20\%. We feel that this procedure
gives reasonable estimates;  since a good understanding of the infrared
region is still lacking, it is difficult to go beyond.

Within our simple model which takes into account the internal
(transverse) motion of the quarks, corrections to the branching ratio
coming from hard gluon exchange between the quark constituents are
small. This confirms the fact that such gluons are not sufficient to
account for the magnitude of large momentum transfer processes (for a
discussion of these problems see \cite{JD}).

The rate of $B\to K^\ast \gamma$ has recently been recalculated,
using various QCD sum rule techniques \cite{ABS,Nar,Ball}. Our
approach yields results most similar to the light-cone sum rules
of ref.~\cite{ABS}; in particular, the form factor $F[O_7]$
scales as $m_B^{-3/2}$. In contrast, it goes as $m_B^{+1/2}$
at maximum recoil $p_\gamma^2 = (m_B-m_V)^2$ in ref.~\cite{Nar}.
We believe that such a scaling is not correct. Using the Gordon
decomposition and the known result $f_B \sim m_B^{-1/2}$, one
obtains on general grounds that $F[O_7]$ scales as $m_B^{-3/2}$
at maximum recoil.

\subsection*{Acknowledgements}
We would like to thank V.~Braun and G.~Kramer for helpful comments
on the manuscript.
\def\etal{et al.}
\gdef\journal#1, #2, #3, #4 { {\sl #1~}{\bf #2}\ (#3)\ #4 }
\def\pr{\journal Phys. Rev., }
\def\prd{\journal Phys. Rev. D, }
\def\prl{\journal Phys. Rev. Lett., }
\def\jmp{\journal J. Math. Phys., }
\def\np{\journal Nucl. Phys., }
\def\pl{\journal Phys. Lett., }

\vspace*{\fill}
\subsection*{Figure captions}
\begin{description}

\item [Fig.\,1:] Leading order contribution from $O_7$, which we use to
determine the rate.

\item [Fig.\,2:] Order $\a_s$ matrix elements of $O_2$ with an
  additional gluon exchange between the internal quark loop and the
  final state $s$-quark. A cross ($\times$) indicates from where
  the photon can be emitted. The dashed line denotes the cut
  generating the absorptive phase.

\item [Fig.\,3:] Order $\a_s$ matrix elements of $O_2$ with an
  additional gluon exchange between the internal quark loop and the
  spectator quark. A cross ($\times$) indicates from where the photon
  can be emitted. The dashed line denotes the cut generating the
  absorptive phase.

\item [Fig.\,4:] Order $\a_s$ corrections to the matrix element of
  $O_7$ with an additional gluon exchange between the $b$- or
  $s$-quark and the spectator.

\end{description}

\newpage
\subsection*{Tables}
\begin{table}[h]
\begin{center}
\begin{tabular}{|l|l|l|}
\hline
$f_B$ [MeV] & $p_F$ [MeV] & $m_{b,eff}$ [GeV] \\  \hline
 150   &  516  &  4.68  \\ \hline
 200   &  643  &  4.53  \\ \hline
 250   &  778  &  4.37  \\ \hline
\end{tabular}
\caption{Wave function parameter $p_F$ and effective
$b$-quark mass $m_{b,eff}$}
\end{center}
\end{table}

\def\bis{\ldots}
\begin{table}[h]
\begin{center}
\begin{tabular}{|l|r|l|l|l|l|}
\hline
$f_B$ & $\rho$~~~~ & R & BR
 & $\acp$ for $\Lambda=1.0$ GeV & $\acp$ for $\Lambda=0.2$ GeV \\
$$[MeV] &  & [\%] & [$10^{-5}$] & [\%] & [\%] \\ \hline
\hline
\multicolumn{6}{|c|}{$B\to K^\ast \gamma$}\\
\hline
150   & ---~~ & 12 &  3.6  &
0.23 \bis 0.76 (0.18 \bis 0.60) & 0.30 \bis 1.0 (0.23 \bis 0.76)  \\ \hline
200   & ---~~ & 14 &  4.3  &
0.24 \bis 0.81 (0.18 \bis 0.60) & 0.32 \bis 1.1 (0.23 \bis 0.77)\\ \hline
250   & ---~~ & 15 &  4.8  &
0.25 \bis 0.85 (0.18 \bis 0.60) & 0.34 \bis 1.1 (0.23 \bis 0.77) \\ \hline
\hline
\multicolumn{6}{|c|}{$B\to \rho \gamma$}\\
\hline
150   & $-$0.3 & 12 &  0.30 
& $-$2.7 \bis $-8.1$ ($-$2.2 \bis $-$6.3) & $-$3.6 \bis $-$11 ($-$2.7 \bis
$-$8.1) \\ \hline
200   &  0.0  & 14 &  0.21 
& $-$4.9 \bis $-$13~ ($-$3.6 \bis $-$9.8) & $-$6.4 \bis $-$18 ($-$4.6 \bis
$-$13)\\ \hline
250   & $+$0.3 & 15 &  0.12 
& $-$10~ \bis $-$24~ ($-$7.2 \bis $-$17~) & $-$14. \bis $-$31 ($-$9.3 \bis
$-$22)\\ \hline
\end{tabular}
\caption{Ratio of exclusive to inclusive decays R,
branching ratio BR, and CP-asymmetry $\acp$
for the decays $B \to K^\ast \gamma$ and  $B \to \rho \gamma$
and for two different cut-off parameters $\Lambda$.
The numerical values correspond to $\eta$ varying in the range
$0.15 \bis 0.5$ and $\alpha_s=0.2$. The weak $\eta$-dependence
of the BR($B \to \rho \gamma$) is not shown.
The numbers in parentheses correspond to taking into account
only the contributions of Fig.~2 in $F[O_2]$.}
\end{center}
\end{table}

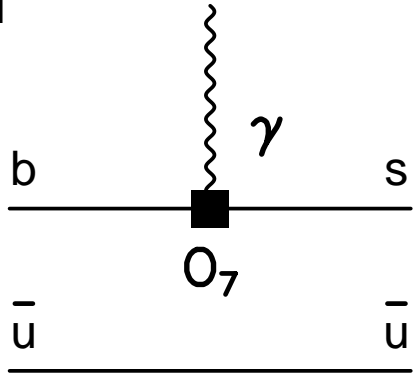
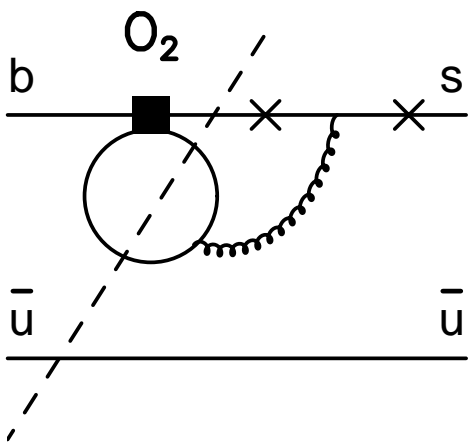
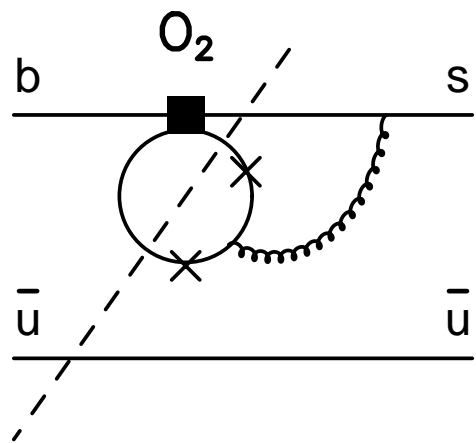
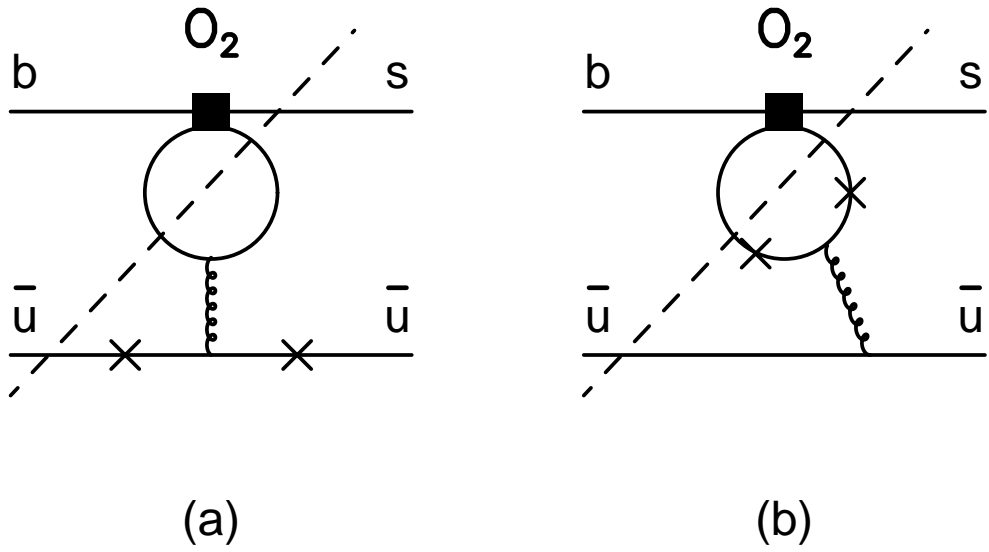
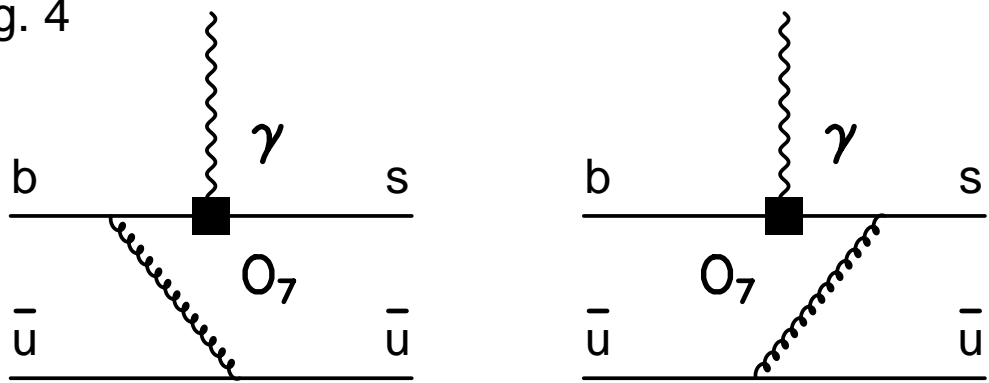
\end{document}